\documentclass[fleqn,usenatbib]{mnras}
\usepackage{graphicx}
\usepackage{newtxtext,newtxmath}
\usepackage[T1]{fontenc}
\usepackage{savesym}
\savesymbol{leftrotoavesymbol{uproot}}
\savesymbol{iint}
\savesymbol{iiint}
\savesymbol{iiiint}
\savesymbol{idotsint}
\savesymbol{Bbbk}
\usepackage{amsmath}	
\usepackage{array}
\usepackage{bm}
\usepackage[dvipsnames,svgnames]{xcolor}
\usepackage{soul}
\usepackage[ruled,lined,linesnumbered]{algorithm2e}
\usepackage{amssymb,physics,esint}
\usepackage{multirow}
\usepackage{booktabs}
\usepackage{multirow}
\usepackage{caption}
\usepackage{lineno}

\DeclareRobustCommand{\VAN}[3]{#2}
\let\VANthebibliography\thebibliography
\def\thebibliography{\DeclareRobustCommand{\VAN}[3]{##3}\VANthebibliography}

\title[Extreme $\gamma$-ray Flares of 3C 454.3 and 3C 279]{Pattern and Origin for the Extreme $\gamma$-ray Flares of 3C 454.3 and 3C 279: An Astrophysical Critical Damper?}

\author[Zhang et al.]{
Haiyun, Zhang$^{1}$
Dahai, Yan$^{2}$\thanks{E-mail: yandahai@ynu.edu.cn}
Jianeng Zhou$^{3}$
Li, Zhang$^{2}$\thanks{E-mail: lizhang@ynu.edu.cn}
Niansheng, Tang$^{1}$\thanks{E-mail: nstang@ynu.edu.cn}
\\
$^{1}$Yunnan Key Laboratory of Statistical Modeling and Data Analysis, Yunnan University, Kunming 650091, People's Republic of China\\
$^{2}$Department of Astronomy, Key Laboratory of Astroparticle Physics of Yunnan Province, Yunnan University, Kunming 650091, People's Republic of China\\
$^{3}$Shanghai Astronomical Observatory, Chinese Academy of Sciences, 80 Nandan Road, Shanghai 200030, People's Republic of China
}

\date{Accepted XXX. Received YYY; in original form ZZZ}

\pubyear{2025}

\begin{document}
\label{firstpage}
\pagerange{\pageref{firstpage}--\pageref{lastpage}}
\maketitle

\begin{abstract}

We apply a Gaussian process method to the extreme $\gamma$-ray flares of 3C 454.3 and 3C 279 to discover the variable patterns and then to investigate the physical origins of the giant flares.
The kernels of stochastically driven damped simple harmonic oscillator (SHO),
the damped random-walk (DRW), and Mat$\acute{\rm e}$rn$-3/2$ are respectively used to describe the adaptive-binning $\gamma$-ray light curves of the two flares.
Our findings show that both the extreme $\gamma$-ray flares of 3C 454.3 and 3C 279 clearly prefer the SHO kernel in the over-damped mode and the Mat$\acute{\rm e}$rn$-3/2$ kernel over the DRW kernel. The resulted SHO and Mat$\acute{\rm e}$rn$-3/2$ power spectral densities (PSDs) are the same for each object, with the index changing from -4 at high frequencies to 0 at low frequencies.
The patterns of the two flares are both approaching the critical damping mode
with the quality factor $Q\approx0.4$ (i.e., the damping ratio $\eta \approx1.25$),
but with slightly different damping timescales.
The characteristic timescale (corresponding to the broken frequency in the PSD) for 3C 454.3 is 2-3 days and 3-5 days for 3C 279.
The variable patterns found here suggest that once the system responds to the energy injection disturbance, 
the release of the energy in the system is finished abruptly.
The obtained timescale provides a constraint on the size of energy dissipation region for each source. 
\end{abstract} 


\section{Introduction} \label{sec:intro}
Jets are powerful collimated plasma flows produced in astrophysical accreting systems. 
Although the production of the relativistic jet for an active galactic nucleus (AGN) is unclear, some clues suggest that the production may be controlled by the spin of the central supermassive black hole (SMBH), the magnetic and material environments near the SMBH \citep[e.g.,][]{2015ASSL..414...45T}.
Blazars are a type of AGN with the jet oriented towards the line of observer's sight.
Blazar jets produce nonthermal radiation from radio to $\gamma$-ray energies, which is characterized by features such as a double-peaked energy spectrum and intense variability.

The variability features of blazar jets mainly involve quasi-periodic oscillations (QPOs) \citep[e.g.,][]{2015ApJ...813L..41A,2018NatCo...9.4599Z,2021ApJ...919...58Z}, 
long-term and short-term stochastic variability \citep{2022ApJ...930..157Z}.
These variability may be related with activities of the central engine \citep[e.g.,][]{2011MNRAS.418L..79T}, 
changes in the physical properties of the jet, for example, the particle acceleration mechanism \citep[e.g.,][]{2013ApJ...765..122Y}, 
or the interaction of the jet with surrounding material \citep[e.g.,][]{2017ApJ...841...61A}.
Possible QPO signals have been reported in many blazars, with several physical explanations proposed \citep[e.g.,][]{2015ApJ...813L..41A,2016NewA...45...32W,2017Ap&SS.362...99W,2018MNRAS.478.3199B,2018ApJ...867...53Y}. 
However, due to challenges in estimating the background noise of AGN variability and the limitations of the Fourier transform-based methods, 
the reliability of most QPOs has been doubted \citep[e.g.,][]{2016MNRAS.461.3145V,2019MNRAS.482.1270C,2021ApJ...907..105Y}.
Gaussian process (GP) method has been introduced to cross-check the reliability of the Blazar QPO in the time domain analysis \citep[e.g.,][]{2020ApJ...895..122C,2021ApJ...907..105Y,2021ApJ...919...58Z,2025MNRAS.537.2380Z}.
It has been shown to be a more accurate method for identifying QPOs in the irregularly sampled time-series than the Fourier transform-based methods. \citep[e.g.,][]{2024MNRAS.531.4181O}.         

We have systematically studied the long-term variability of nonthermal jet radiation by using the GP method and found that their variability mode adheres to the damped random-walk (DRW) model.
Furthermore, the long-term jet variability may be related to thermal instabilities in the accretion disk \citep{2022ApJ...930..157Z,2023ApJ...944..103Z}.
On the other hand, the high-confidence QPO signal detected in non-blazar PKS 0521-36 \citep{2021ApJ...919...58Z} also supports the scenario that the long-term jet variability is related to activities of the AGN central engine.

The short-term variability of blazar jet usually reflects the local condition of the jet where the emissions are produced. 
Rapid $\gamma$-ray flares allow us to probe the physical processes working
in the innermost regions of jets \citep[e.g.,][]{2016ApJ...824L..20A,2018ApJ...854L..26S}.
Minute-timescale variability detected at $\gamma$-ray energies indicates a very compact emission region, 
where energy dissipation and particle acceleration take place in the extreme physical conditions.
Although blazar flares are often proposed to be driven by shocks \citep[e.g.,][]{2010ApJ...711..445B}, 
the shock model has difficulties explaining the extreme flares \citep[e.g.,][]{2015MNRAS.450..183S}.
Recent studies show that magnetic reconnection is a promising physical mechanism for explaining blazar bright flares \citep[e.g.,][]{2020MNRAS.494.1817D,2022Natur.609..265J}.
It allows for fast energy release and nonthermal particle acceleration 
in high-energy astrophysics \citep{2016MNRAS.462...48S,2020PhPl...27h0501G,2015MNRAS.450..183S,2018ApJ...864..164Y}.

In our previous work \citep{2022ApJ...930..157Z}, 
we observed that the fast and bright $\gamma$-ray flares 
were not well described by the DRW model. 
This could imply that the long-term jet variability and giant flares have distinct variability patterns.
In this work, we focus on the brightest flare of 3C 454.3 detected by the Large Area Telescope on the Fermi Gamma-ray Space Telescope (Fermi-LAT) in November 2010 and the flare of 3C 279 detected in January 2018.
The goal is to identify the variability patterns and the mechanism of the bright outbursts in GeV $\gamma$-ray energies.
This paper is structured as follows: Section \ref{sec:method} outlines the method and models that we used to process and analyze the data.
Section \ref{sec:results} displays the fitting results for 3C 454.3 and 3C 279 flares by three different kernels.
Section \ref{sec:sampling method} examines the impact of parameter autocorrelation and sampling methods on the results.
In Section \ref{sec:discussion}, we discuss the implications of our results with respect to the jet physics of 3C 454.3 and 3C 279 in the extremely flaring states.
Finally, a summary is given in Section \ref{sec:summary}.

\section{Method and data processing} \label{sec:method}

Fermi-LAT is a pair production detector with a 2.4 sr wide field-of-view (FoV), operating in survey mode and covering energies from tens of MeV to over hundreds of GeV. In analysis, 0.1-300 GeV Pass 8 data within a region of interest (ROI) of $20^{\circ} \times 20^{\circ}$ around 3C 454.3 and 3C 279 are selected. Standard procedure with FermiTools (v2.0.8) is used to performed likelihood analysis. Corresponding instrument response function (IRF) is P8\_SOURCE\_V3. During analysis, the background is modeled by sources cataloged in LAT 10-year source catalog (4FGL-DR2; \cite{2020ApJS..247...33A}) and the diffuse models (Galactic diffuse template gll\_iem\_v07.fits and extragalactic isotropic diffuse template iso\_P8R3\_SOURCE\_V3\_v1.txt).

The adaptive binning method \citep{2012A&A...544A...6L} is used to construct light curves, which can create unevenly binned light curve with constant uncertainty. We adopt a 12\% uncertainty level following the strategy proposed and validated by \cite{2012A&A...544A...6L}. This threshold falls within a practically motivated range and provides a reasonable balance between time resolution and statistical robustness.
This construction can reveal more information in the light curve than with the fixed-binning method. Likelihood analysis on a wider time range is applied for 3C 454.3 (MJD 55400 to MJD 55670) and 3C 279 (MJD 58116 to MJD 58190) to derive a model as global description of sources in the ROI. Spectral parameters in this model are used to perform likelihood analysis in each bin once adaptive bins are given. Noticeably, spectral shapes of background sources are fixed in light curve calculation.

For extracting variability features from light curves, we use a powerful data processing method, i.e., GP method \citep{2023ARA&A..61..329A}.
It is a statistical model based on probability theory and Bayesian analysis techniques.
A GP can be completely determined by the mean function and the covariance (kernel) function.
The mean function is generally set to the mean value of the time series.

In the astronomical field, DRW and the stochastically driven damped simple harmonic oscillator (SHO) models are commonly used kernels \citep[e.g.][]{2021Sci...373..789B,2022ApJ...936..132Y,2022ApJ...930..157Z}.
The DRW model is typically used to capture the random fluctuations in AGN light curves, 
while the SHO model can describe periodic behaviors.
The details on the introduction and applications of both models are described in the earlier papers \citep{2021ApJ...919...58Z,2022ApJ...930..157Z,2023ApJ...944..103Z,2024ApJ...971...26T}.

Here, we emphasize the PSDs in theory predicted by the two kernels.
The PSD of DRW model is calculated using the following formula:
\begin{equation}
    S(\omega)= \sqrt\frac{8}{\pi}\frac{\sigma_{\rm DRW}^{2}\tau_{\rm DRW}}{1+\tau_{\rm DRW}^2 \omega^2},\label{DRW_PSD}
\end{equation}
where $\tau_{\rm DRW}$ and $\sigma_{\rm DRW}$ are damping timescale and the standard deviation of variability respectively.
It follows a broken power-law form, with a constant index of zero at low frequencies, 
and a form of $\nu^{-2}$ ($\nu=\omega/2\pi$ is frequency) decay at high frequencies \citep{2009ApJ...698..895K}. The broken frequency is $\nu_{b}=1/(2\pi\tau_{\rm DRW})$.
The PSD of SHO model is:
\begin{equation}\label{eq3}
S(\omega)=\sqrt{\frac{2}{\pi}}\frac{S_{0}\omega_{0}^{4}}{(\omega^{2}-\omega_{0}^{2})^{2}+\omega^{2}\omega_{0}^{2}/Q^{2}}\;,
\end{equation}
where $S_0=S(\omega_0)$ and $Q$ is quality factor.
It is more complex compared to DRW PSD. 
In the over-damped mode ($Q<0.5$), it behaves similarly to the DRW PSD at low frequencies, while at high frequencies, the index can steepen to $-4$ \citep{2019PASP..131f3001M}.
In theory, there are two timescales (corresponding to two broken frequencies in the SHO PSD, i.e., $\sim 1/\nu_{b}$), which are denoted as \citep{2017MNRAS.470.3027K}:
\begin{equation}\label{eq:t_flat}
t_{\rm flat}={\rm max}\ (\frac{2\pi}{\omega_{0}},\frac{2\pi}{\omega_{0}Q}\sqrt{1-2Q^{2}})\;,
\end{equation}
\begin{equation}\label{eq:t_steep}
t_{\rm steep}={\rm min}\ (\frac{2\pi}{\omega_{0}},\frac{2\pi Q}{\omega_{0}\sqrt{1-2Q^{2}}})\;.
\end{equation}
At $t_{\rm flat}$, the PSD distribution index changes from $-2$ to $0$, whereas at $t_{\rm steep}$, it changes from $-4$ to $-2$.
If SHO is in the under-damped mode ($Q>0.5$), a peak will appear in the PSD.
The critically damped mode ($Q=0.5$) of SHO is the fastest mode to lose the correlation between the data \citep{2017MNRAS.470.3027K,2019PASP..131f3001M}.
In physics, the critically damped mode refers to a specific type of damping in a dynamic system which reaches equilibrium as quickly as possible.
 
Besides the commonly used DRW and SHO kernels,
here we introduced the Mat$\acute{\rm e}$rn class kernels.
It is characterized by a parameter $l$ that controls the smoothness of the function.
For simplicity, the value of $l$ is usually chosen to be a half-integer.
Taking $l=3/2$ (Mat$\acute{\rm e}$rn$-3/2$), it provides a good balance between smoothness and flexibility \citep{2006gpml.book.....R}, whose function is
\begin{equation}
  k (\tau)=\sigma^{2}(1+\frac{\sqrt{3}\tau}{\rho}){\rm exp}(-\frac{\sqrt{3}\tau}{\rho})\ ,
    \label{maternkernel}
\end{equation}
where $\rho$ corresponds to a characteristic timescale, and $\sigma$ represents the term of the amplitude.
The PSD of Mat$\acute{\rm e}$rn$-3/2$ is a broken power-law form, with the index changing from $-4$ to $0$ from high frequencies to low frequencies. The broken frequency is $\nu_{b}=1/(2\pi\rho)$. Actually, Mat$\acute{\rm e}$rn$-3/2$ can be approximated by the critically damped mode of the SHO model \citep{2006gpml.book.....R,2017AJ....154..220F}.

We employ the {\it Celerite} package \citep{2017AJ....154..220F} to implement the GP method, utilizing, e.g., maximum likelihood estimation and MCMC sampling technique \citep[emcee\footnote{\url{https://github.com/dfm/emcee}},][]{2013PASP..125..306F} in data processing.
We assume uniform priors on the natural logarithm (ln) of each parameter and determine the initial values of these parameters for MCMC sampling using a maximum likelihood estimate executed 100 times.
The MCMC sampler runs for 50,000 iterations with 32 parallel walkers. The ﬁrst 20,000 steps are taken as burn-in. 
The final 30,000 MCMC samples were used to construct the posterior distributions of the parameters and the PSDs.

\section{results} \label{sec:results}

\begin{figure*}
    \centering
    {\includegraphics[width=1\linewidth]{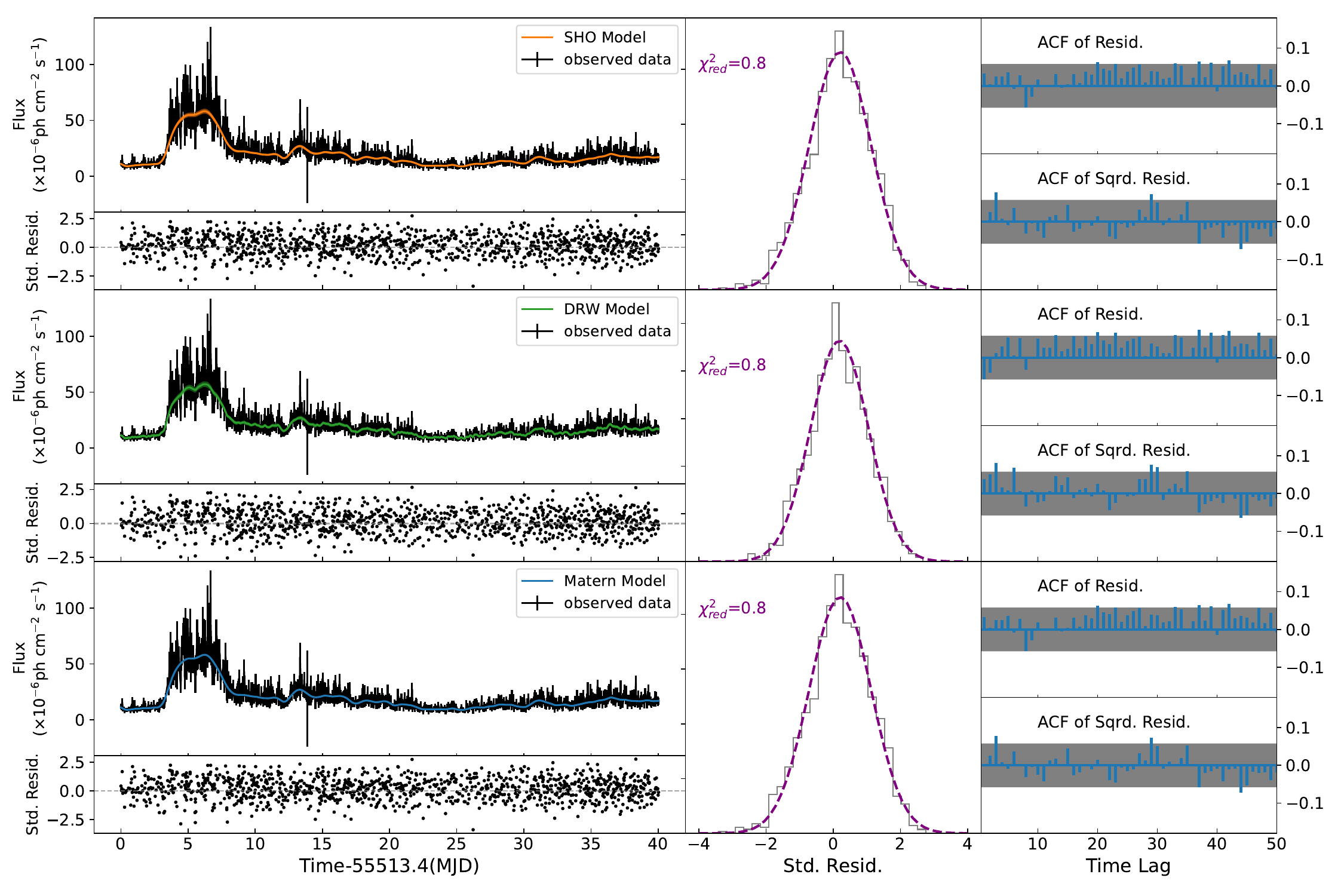}}
    \caption{Fitting results of 3C 454.3 flare. 
    The upper, middle, and lower rows displaying the fittings for the SHO, DRW and Mat$\acute{\rm e}$rn$-3/2$ models, respectively.
    For each row, we give the LAT observed data (the top-left panel), the modeled light curves with model's predictive uncertainty (orange/green/blue line and the shaded region in the top-left panel) and the standardized residuals (the bottom-left panel) in the left column. 
    In the middle column, we show the probability density of standardized residuals (gray histogram) as well as the best-fit normal distribution (purple dotted line) with the best-fit parameters lised in Table~\ref{tab:prior}. The purple annotation in the panel indicates the reduced chi-squared value used to assess the goodness of fit.
    The ACFs of residuals and the ACFs of squared residuals, along with the 95$\%$ confidence interval for white noise (the gray region) are shown in the right column. 
 \label{fig:3C454celerite fit}}
\end{figure*}

\begin{figure*}
    \centering
    {\includegraphics[width=1\linewidth]{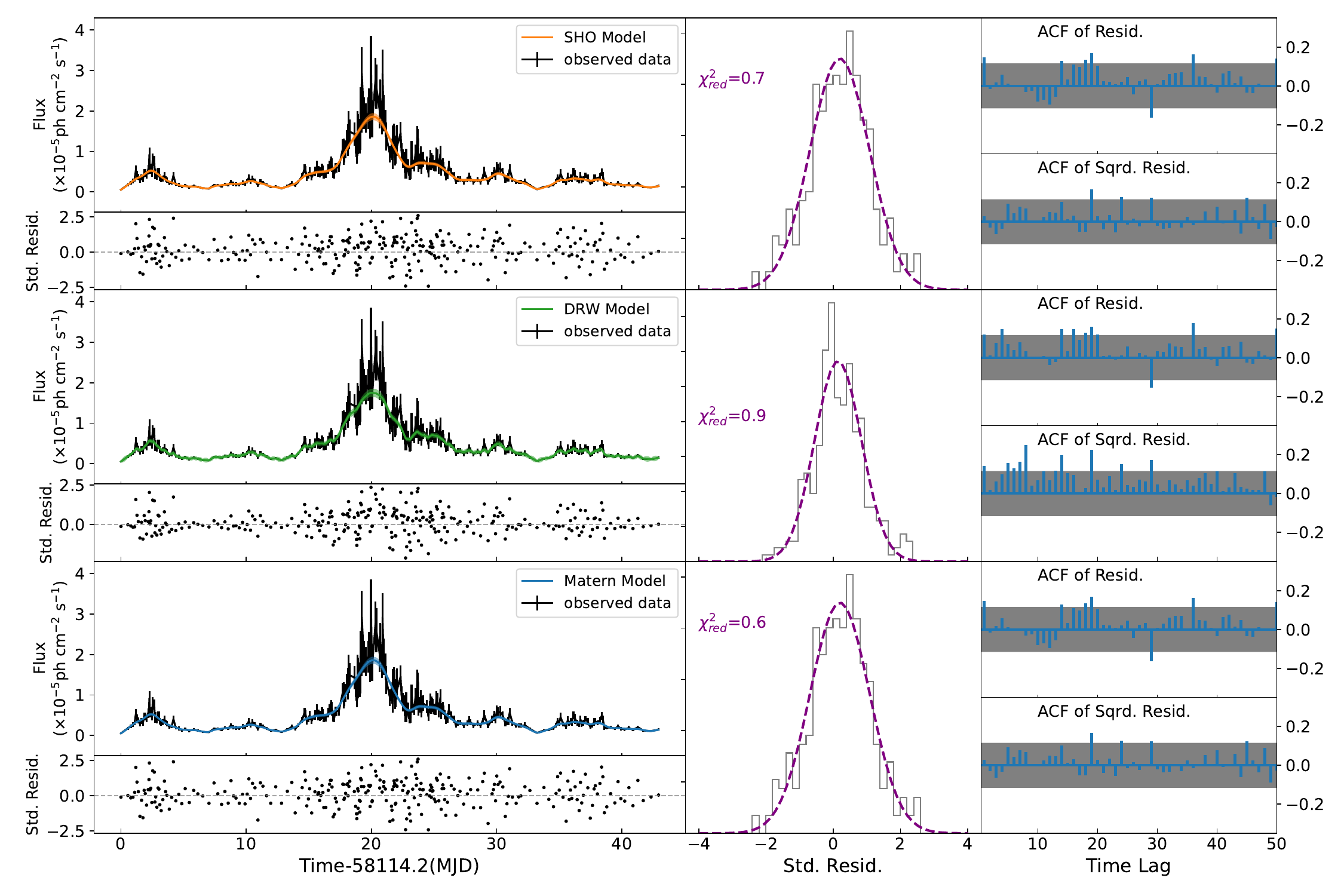}}
    \caption{Fitting results of 3C 279 flare. 
    The symbols and lines are the same as those in Figure~\ref{fig:3C454celerite fit}.
\label{fig:3C279celerite fit}}
\end{figure*}

3C 454.3 and 3C 279 are bright flat spectrum radio quasars (FSRQs) with redshift of 0.86 and 0.54 respectively.
They show intense activity in GeV $\gamma$-rays. 
The extreme flare (MJD 55513-MJD 55553) of 3C 454.3 and the flare (MJD 58116-MJD 58159) of 3C 279 detected by Fermi-LAT are analyzed here.
Light curves of the two flares are constructed using the adaptive-binning method, producing 1141 data points for 3C 454.3 flare and 287 data points for 3C 279 flare (the black points in the top-left panels of each row in Figure~\ref{fig:3C454celerite fit} and Figure~\ref{fig:3C279celerite fit}).

To accurately grasp the characteristics of the flares, we fit their light curves with three distinct kernels, i.e., DRW, SHO and Mat$\acute{\rm e}$rn$-3/2$ separately.
The priors of the parameters for each model are listed in Table~\ref{tab:prior}.
Figure~\ref{fig:3C454celerite fit} (for 3C 454.3) and Figure~\ref{fig:3C279celerite fit} (for 3C 279) show the fitting results, with the upper, middle and lower rows displaying the fittings for the SHO, DRW and Mat$\acute{\rm e}$rn$-3/2$ models, respectively. 
The fitted light curves with their uncertainties in the figures are constructed by the median values of all hyperparameters sampled by the MCMC.
According to the goodness-of-fit criteria described in \cite{2022ApJ...930..157Z}, there are no significant differences among the fitting plots of the three models, making it difficult to determine which model is the best. 
The corrected Akaike information criterion ($\rm AIC_{\rm C}$) 
is therefore used to perform the model selection.
For both 3C 454.3 and 3C 279, the $\rm AIC_{\rm C}$ values for SHO and Mat$\acute{\rm e}$rn$-3/2$ models are very close to each other, and significantly smaller than that of DRW model ($\Delta \rm AIC_{\rm C}\textgreater10$) as shown in Table~\ref{tab:prior}.
This suggests that the SHO and Mat$\acute{\rm e}$rn$-3/2$ models are more appropriate for the data of the flares.

\begin{figure*}
    \centering
    \begin{minipage}{0.45\textwidth}
      \centering
      \includegraphics[width=1\linewidth]{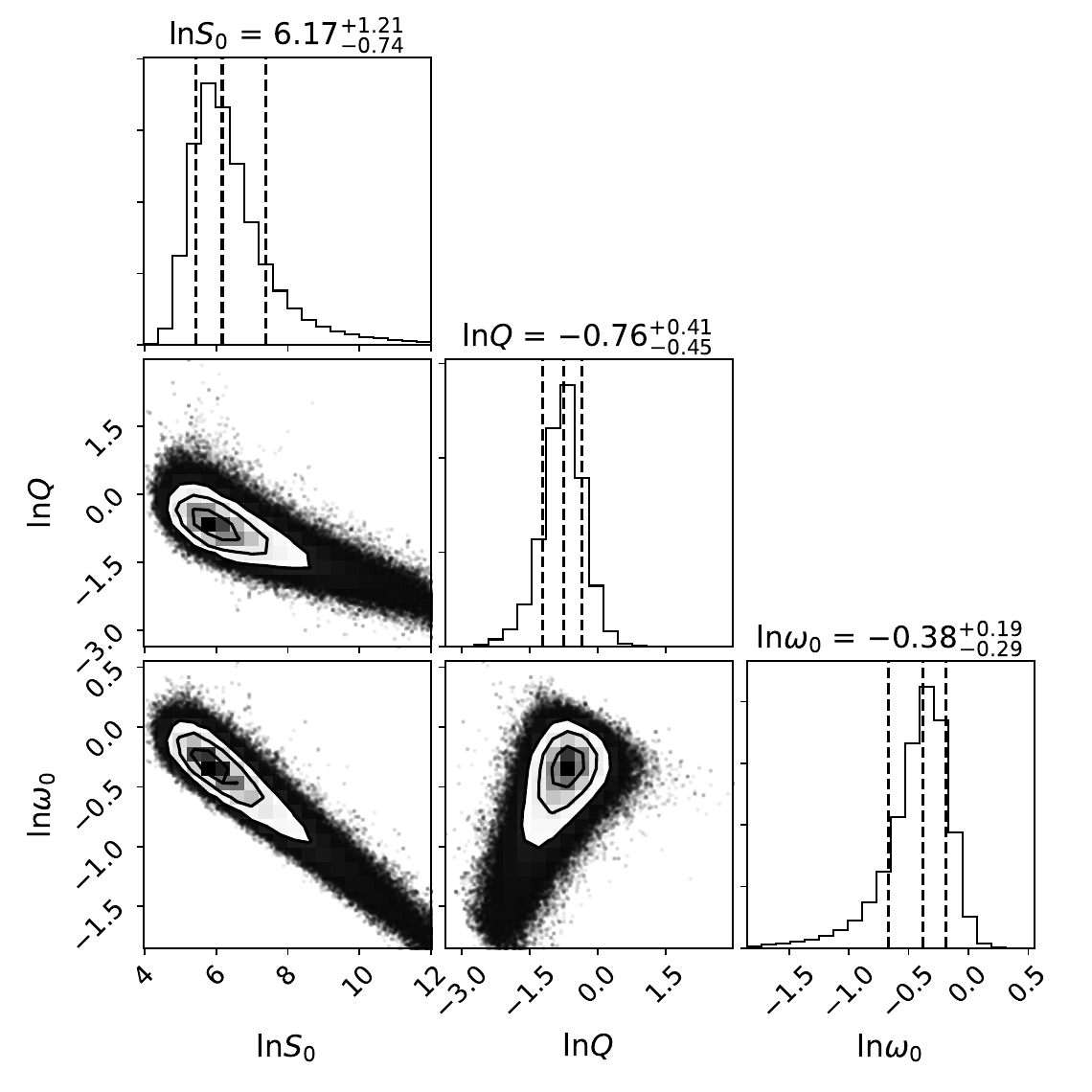}
      \label{fig:3C454SHO}
    \end{minipage} \hfill
    \begin{minipage}{0.45\textwidth}
      \centering
      \includegraphics[width=1\linewidth]{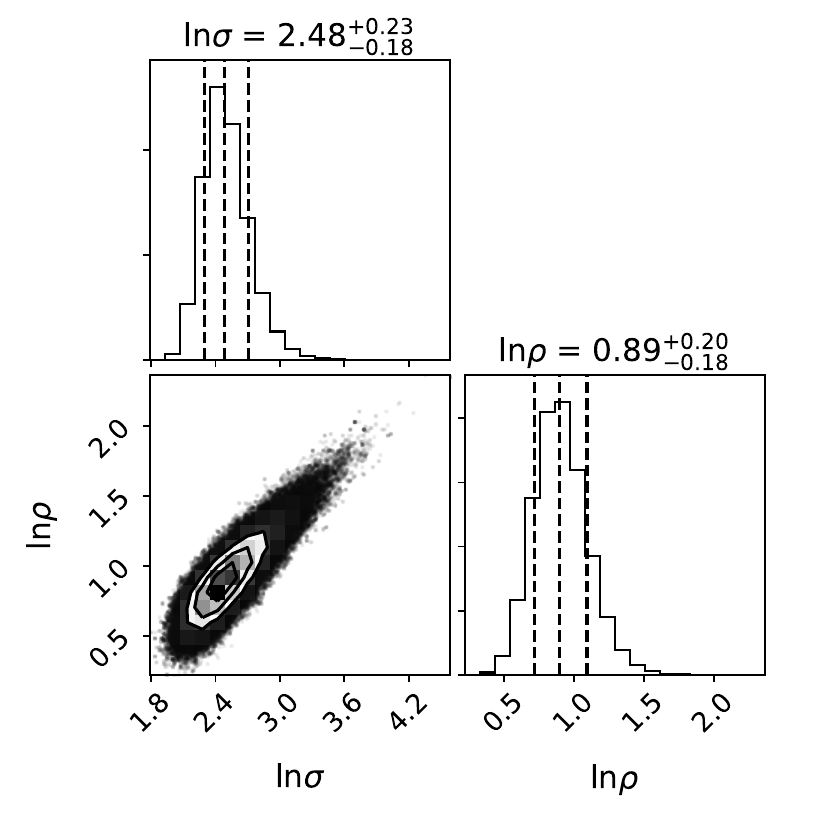}
      \label{fig:3C454M}
    \end{minipage} \hfill
    \begin{minipage}{0.45\textwidth}
      \centering
      \includegraphics[width=1\linewidth]{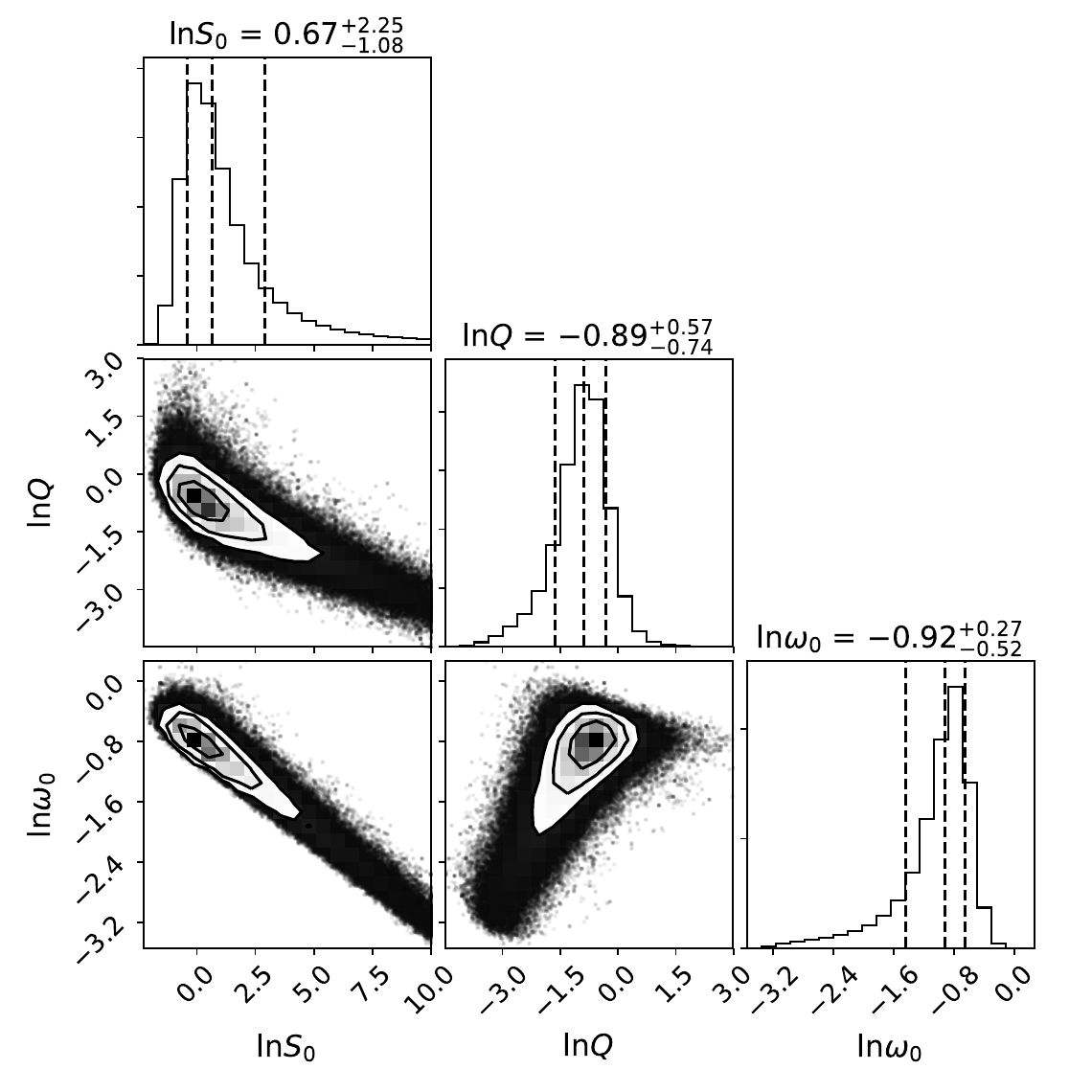}
      \label{fig:3C279SHO}
    \end{minipage} \hfill
    \begin{minipage}{0.45\textwidth}
      \centering
      \includegraphics[width=1\linewidth]{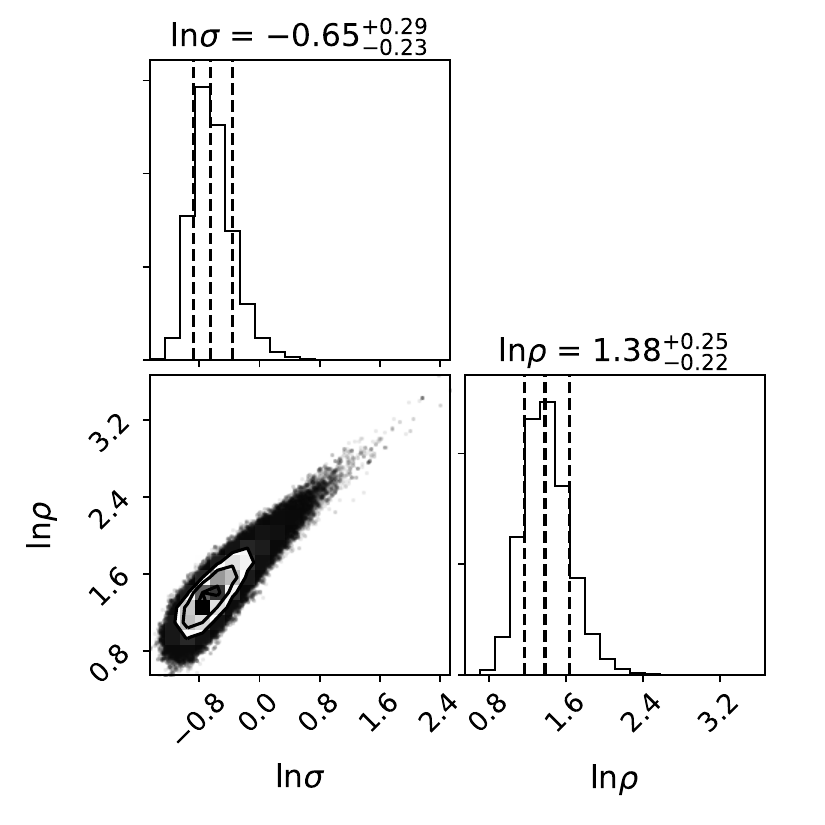}
      \label{fig:3C279M}
    \end{minipage} 
    \caption{The posterior probability density distribution of the SHO (the left column) and Mat$\acute{\rm e}$rn$-3/2$ (the right  column) parameters for the 3C 454.3 flare (the top row) and 3C 279 flare (the bottom row). 
    The vertical dotted lines represent the median value and 68$\%$ confidence intervals of the distribution of the parameters.   
    \label{fig:param}}
\end{figure*}

We then show the posterior probability density distribution of the SHO and Mat$\acute{\rm e}$rn$-3/2$ parameters (under the natural logarithm) for the two giant flares in Figure~\ref{fig:param}. 
The parameters for both models are effectively constrained.
The parameter values for the three models are listed in Table~\ref{tab:Fitting results}. 

\begin{table*}
\caption{Priors and Fitting Information for Each Model}
\resizebox{1\textwidth}{!}{
\begin{tabular}{ccccccccc}
\hline\hline
Source & Model & Parameter & Prior & Autocorrelation & $\rm AIC_{\rm C}$ &  $\ln Z$ & $\mu_{\rm fit}$ & $\sigma_{\rm fit}$\\
\hline
\multirow{7}{*}{3C 454.3} 
  & \multirow{3}{*}{SHO} 
    & ln($S_{0}$)      & Log-uniform(-20, 12) & 46 & \multirow{3}{*}{1261} & \multirow{3}{*}{-3264} & \multirow{3}{*}{$0.2\pm0.03$} & \multirow{3}{*}{$0.9\pm0.02$}\\
  &                   & ln($Q$)              & Log-uniform(-15, 15) & 45 & & & &\\
  &                   & ln($\omega_{0}$)     & Log-uniform(-2, 5) & 46 & & & &\\
\cmidrule(lr){2-9}
  & \multirow{2}{*}{DRW} 
    & ln(a)           & Log-uniform(-7, 7) & 36 & \multirow{2}{*}{1292} & \multirow{2}{*}{-3277} & \multirow{2}{*}{$0.2\pm0.03$} & \multirow{2}{*}{$0.9\pm0.02$}\\
  &                   & ln(c)              & Log-uniform(-6, 7) & 36 & & & &\\
\cmidrule(lr){2-9}
  & \multirow{2}{*}{Mat$\acute{\rm e}$rn-3/2} 
    & ln($\sigma$)    & Log-uniform(-5, 5) & 33 & \multirow{2}{*}{1259} & \multirow{2}{*}{-3261} & \multirow{2}{*}{$0.2\pm0.03$} & \multirow{2}{*}{$0.9\pm0.02$}\\
  &                   & ln($\rho$)         & Log-uniform(-5, 5) & 33 & & & &\\
\hline
\multirow{7}{*}{3C 279} 
  & \multirow{3}{*}{SHO} 
    & ln($S_{0}$)      & Log-uniform(-20, 10) & 47 & \multirow{3}{*}{-276} & \multirow{3}{*}{132} & \multirow{3}{*}{$0.2\pm0.05$} & \multirow{3}{*}{$0.9\pm0.04$}\\
  &                   & ln($Q$)              & Log-uniform(-5, 3) & 45 & & & &\\
  &                   & ln($\omega_{0}$)     & Log-uniform(-10, 10) & 47 & & & &\\
\cmidrule(lr){2-9}
  & \multirow{2}{*}{DRW} 
    & ln(a)           & Log-uniform(-11, 4) & 33 & \multirow{2}{*}{-251} & \multirow{2}{*}{123} & \multirow{2}{*}{$0.1\pm0.04$} & \multirow{2}{*}{$0.7\pm0.04$}\\
  &                   & ln(c)              & Log-uniform(-6, 5) & 32 & & & &\\
\cmidrule(lr){2-9}
  & \multirow{2}{*}{Mat$\acute{\rm e}$rn-3/2} 
    & ln($\sigma$)    & Log-uniform(-5, 5) & 39 & \multirow{2}{*}{-278} & \multirow{2}{*}{135} & \multirow{2}{*}{$0.2\pm0.05$} & \multirow{2}{*}{$0.9\pm0.04$}\\
  &                   & ln($\rho$)         & Log-uniform(-5, 5) & 39 & & & &\\
\hline
\end{tabular}
}
\label{tab:prior}
\vspace{0.5em}
\footnotesize
\parbox{\textwidth}{\small Note: $a = 2\sigma_{\rm DRW}^{2}$, $c = 1/\tau_{\rm DRW}$. Autocorrelation refers to the autocorrelation of parameters computed during MCMC sampling. $\ln Z$ is the log Bayesian evidence calculated for each model in the nested sampling method. The priors are in the natural logarithm. $\mu_{\rm fit}$ (approaches 0) and $\sigma_{\rm fit}$ (approaches 1) are the best-fit parameters of the normal distribution for the standardized residuals.}
\end{table*}

\begin{figure*}
    \centering
    {\includegraphics[width=0.45\linewidth]{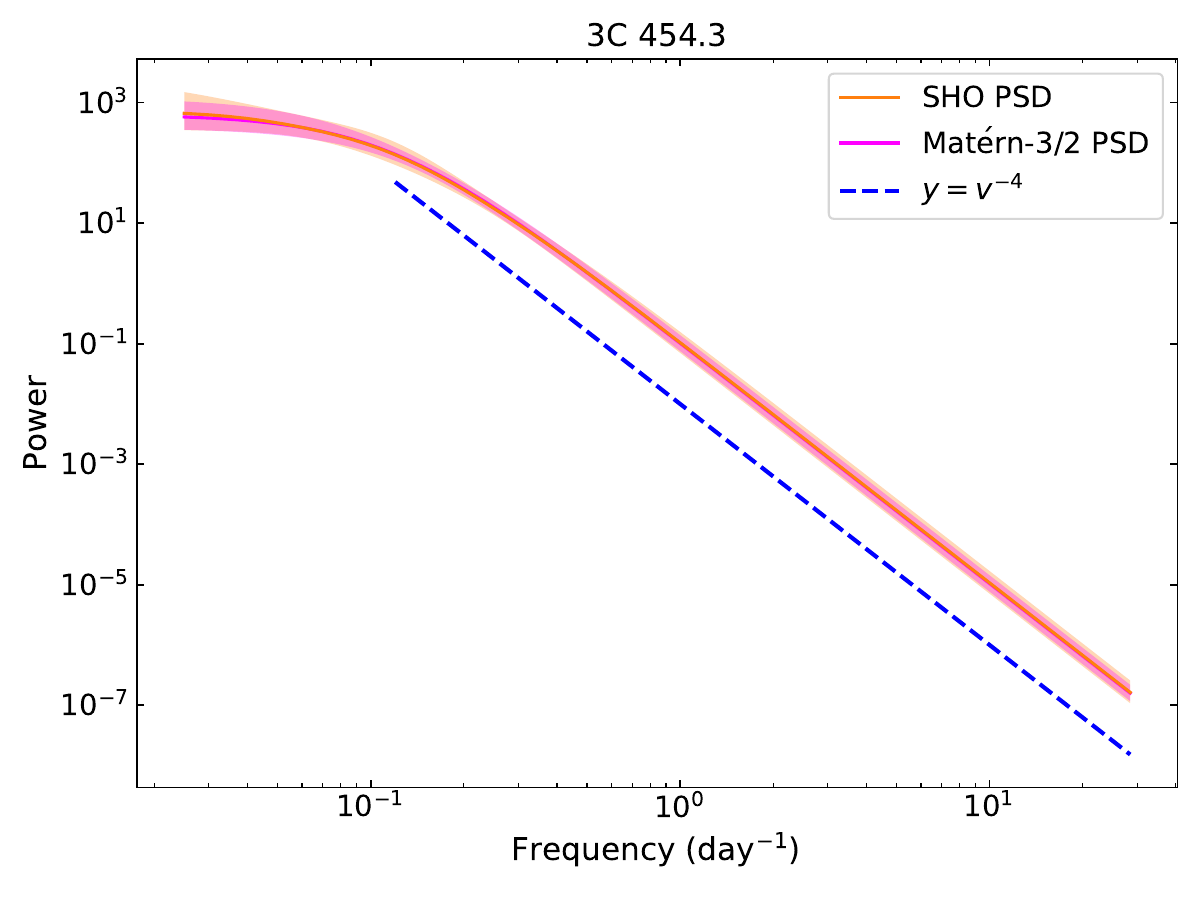}}
    {\includegraphics[width=0.45\linewidth]{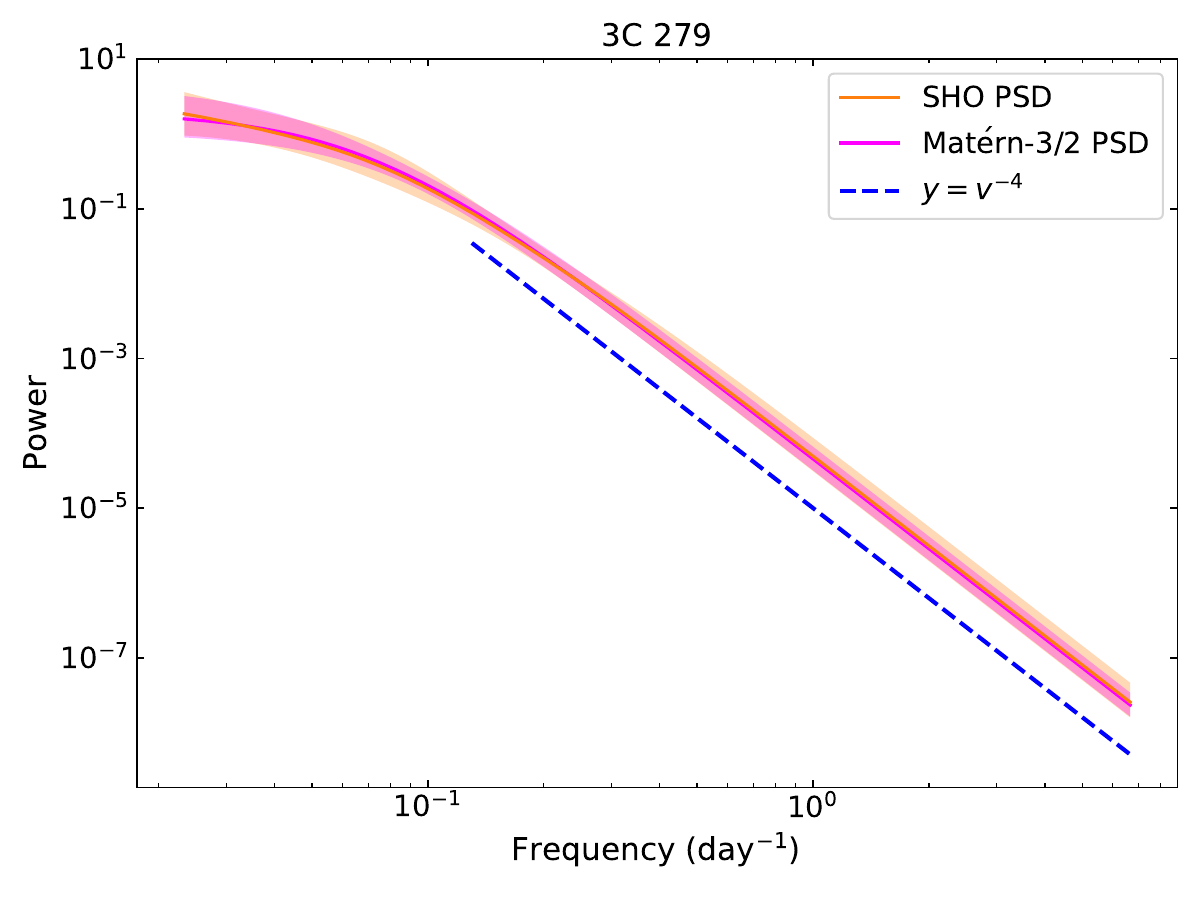}}
    \caption{ PSDs of 3C 454.3 flare (left) and 3C 279 flare (right) constructed from the modeling results of the SHO and Mat$\acute{\rm e}$rn$-3/2$ models. The dashed blue line is a reference line with a power-law index of $-4$. 
\label{fig:psd}} 
\end{figure*}

For the two flares, the SHO model is constrained in the over-damped mode (Q$\textless$0.5).
Interestingly, it is very close to the critically damped mode ($Q=0.5$) 
with the medium value of $Q\approx0.4$.
In this case, the SHO PSD is similar to that of Mat$\acute{\rm e}$rn$-3/2$, with the distribution index changing directly from $-4$ to $0$ (see Figure~\ref{fig:psd}). 
As seen in the power spectrum, the PSDs derived from the two models almost entirely overlap.
In the Mat$\acute{\rm e}$rn$-3/2$ mode, we get the timescales $\rho=2-3$ ($2.4^{+0.6}_{-0.3}$) days for 3C 454.3 and $\rho=3-5$ ($4.0^{+1.0}_{-0.8}$) days for 3C 279.
 In SHO mode, $t_{\rm flat}/2\pi$ are calculated as $2.3^{+2.0}_{-1.6}$ days for 3C 454.3 and $5.0^{+6.0}_{-4.3}$ days for 3C 279, which have great uncertainties and are consistent with the value of $\rho$ within the error range.
$t_{\rm steep}/2\pi$ is $0.9^{+0.8}_{-0.6}$ days and $1.3^{+1.5}_{-1.1}$ days for 3C 454.3 and 3C 279 respectively.
For each source, $t_{\rm steep}/2\pi$ is within the error range of $t_{\rm flat}/2\pi$.

\begin{table*}
\renewcommand{\arraystretch}{1.5}
  \centering
  \caption{Posterior parameters of the models}
  \label{tab:Fitting results}
   \resizebox{1\textwidth}{!}{  
  \centering
  \begin{tabular}{ccccccccc}
  \hline\hline    
  Name & Time period & \multicolumn{3}{c}{SHO Parameters} & \multicolumn{2}{c}{Mat$\acute{\rm e}$rn$-3/2$ Parameters} & \multicolumn{2}{c}{DRW Parameters} \\
  \cmidrule(r){3-5} \cmidrule(r){6-7} \cmidrule(r){8-9} \\
  \addlinespace[-13pt]
   & & ln $S_{0}$ & ln $Q$ & ln $\omega_{0}$ & ln $\sigma$ &
ln $\rho$ & ln $\sigma_{\rm DRW}$ & ln $\tau_{\rm DRW}$ \\
  (1) & (2) & (3) & (4) & (5) & (6) & (7) & (8) & (9) \\
  \hline
  3C 454.3 & MJD 55513-55553 & $6.17^{+1.21}_{-0.74}$ & $-0.76^{+0.41}_{-0.45}$ & $-0.38^{+0.19}_{-0.29}$ & $2.48^{+0.23}_{-0.18}$ & $0.89^{+0.20}_{-0.17}$ & $2.23^{+0.51}_{-0.31}$ & $2.55^{+1.03}_{-0.64}$   \\
3C 279 & MJD 58116-58159 & $0.67^{+2.25}_{-1.08}$ & $-0.89^{+0.57}_{-0.74}$ & $-0.92^{+0.27}_{-0.52}$ & $-0.65^{+0.29}_{-0.23}$ & $1.38^{+0.25}_{-0.22}$ & $-0.78^{+0.66}_{-0.43}$ & $3.44^{+1.33}_{-0.88}$   \\
  \hline
  \end{tabular}
  }
  \parbox{\textwidth}{\small Note: 
(1) source name, (2) time period of the flare, (3)(4)(5) posterior parameters of modeling light curves with SHO model, (6)(7) posterior parameters of Mat$\acute{\rm e}$rn$-3/2$ model, (8)(9) posterior parameters of DRW model. The uncertainties of each model parameter represent $1\sigma$ confidence interval. $\rho$ and $\tau_{\rm DRW}$ is in the unit of day, and $\omega_{0}$ is in the unit of $\rm rad\cdot day^{-1}$.
}
\end{table*}

\section{Impact of Parameter Autocorrelation and Sampling Methods on Inference Results}  \label{sec:sampling method}

In the previous section, our results were based on all MCMC samples after discarding the initial 20,000 steps, without accounting for autocorrelation of the parameters.
In this section, we evaluate the impact of parameter autocorrelation by applying a thinning procedure and comparing the results with those obtained previously.
We further investigate how different sampling methods (nested sampling and MCMC sampling) affect our inference results.

We estimate the autocorrelation length $\tau$ for each parameter in three models using the method implemented in the emcee package. 
This estimate indicates how many steps the sampling chain needs to be spaced apart to obtain approximately independent samples.
The specific $\tau$ values of the parameters for three models are listed in the Table~\ref{tab:prior}.
We adopted the maximum autocorrelation length among all parameters in each model as the basis for thinning the chain.
That is, we retained one sample every $\tau$ steps from the post-burn-in portion of the chains to construct the posterior sample set.
The thinned posterior samples were then used to calculate PSDs, corner plots and fits of the GP model.
For both sources (3C 454.3 and 3C 279), the fitting performance, PSDs and parameters of three models are almost identical with our previous results. We show the posterior parameters of the SHO as an example in Appendix~\ref{sec:appendix_figs} (the top panel of Figure~\ref{fig:param-sampling}). 

In addition to the MCMC approach, we perform GP inference using the nested sampling method, as implemented in the dynesty package. 
We use the same GP models and prior ranges in both sampling methods to ensure consistency. For nested sampling, we adopt the dynamic nested sampler with a conservative convergence criterion. Posterior samples are obtained via importance resampling and then used to compute parameter estimates, PSDs, and GP-predicted light curves.
We find that the inference results obtained using nested sampling are highly consistent with those from MCMC sampling (distribution of SHO parameters as an example shown in the bottom panel of Figure~\ref{fig:param-sampling} in Appendix~\ref{sec:appendix_figs}), indicating that our results are robust with respect to the choice of sampling algorithm.
The logarithmic Bayesian evidence ($\ln Z$) for the three models is calculated and listed in Table~\ref{tab:prior}. In terms of the logarithmic Bayes factor between the models (i.e., $\ln Z_{i}-\ln Z_{j}$), both the SHO and Mat$\acute{\rm e}$rn$-3/2$ models are significantly better than the DRW model ($\ln Z_{\rm Mat\acute{\rm e}rn-3/2}-\ln Z_{\rm DRW}\approx 16(12);\ln Z_{\rm SHO}-\ln Z_{\rm DRW}\approx 13(9)$). However, there is no strong evidence to determine whether the SHO or the  Mat$\acute{\rm e}$rn$-3/2$ model provides a better fit ($\ln Z_{\rm Mat\acute{\rm e}rn-3/2}-\ln Z_{\rm SHO}\approx 3$), which is consistent with the conclusion suggested by the $\rm AIC_{C}$. 

In summary, the parameter autocorrelation in the MCMC method, as well as the choice of sampling method, does not affect our results.

\section{Discussion} \label{sec:discussion}
We have carried out a temporal analysis for the extreme $\gamma$-ray flares (detected by Fermi-LAT) of 3C 454.3 and 3C 279 by the GP method.
SHO, DRW, and Mat$\acute{\rm e}$rn$-3/2$ kernerls are used in the GP analysis, to model the two flares.
We found that the light curves of both the flares can be successfully described by SHO and Mat$\acute{\rm e}$rn$-3/2$ models, with SHO being constrained in the over-damped mode and approaching the critically damped mode ($Q\approx0.4$, i.e., the damping ratio $\eta =1/(2Q)\approx1.25$).
There is no strong evidence indicating which of the two models is better.
The characteristic timescale is 2-3 days for 3C 454.3 and 3-5 days for 3C 279. 

Our previous works showed that long-term jet variability follows the DRW pattern, 
and the damping timescale is associated with the thermal instability timescale of accretion disk \citep{2022ApJ...930..157Z,2023ApJ...944..103Z}.
While for the short-term variability in this work, 
the variability pattern is SHO or Mat$\acute{\rm e}$rn$-3/2$, indicating a different origin.
Based on the variability pattern of the extreme $\gamma$-ray flares, 
we establish a physically motivated scenario to interpret the results of the flare variability in 3C 454.3 and 3C 279.
In this scenario, a compact emitting region (or "jet blob") within the relativistic jet is disturbed by a sudden injection of energy, such as from magnetic reconnection or internal shocks.
The system may not respond instantly; instead, it reacts over a responding/retarding timescale during which internal physical conditions adjust to the disturbance. After this timescale, the energy injection leads to significant structural or radiative changes in the blob, temporarily driving the system out of equilibrium.
Subsequently, the dissipation processes, such as radiative cooling and turbulent dissipation, gradually reduce excess energy, allowing the system to return to equilibrium over a relaxation or dissipation timescale \citep{1943RvMP...15....1C,1945RvMP...17..323W}.

We interpret the results of 3C 454.3 and 3C 279 flares in this physical scenario.
As the Mat$\acute{\rm e}$rn$-3/2$ kernel can be considered as a special case (approaches the critically damped regime of SHO) of the more general SHO model \citep{2006gpml.book.....R,2017AJ....154..220F}, we adopt SHO model for physical interpretation. 
In theory, the PSD of SHO model in over-damped mode has two characteristic frequencies, 
i.e., two corresponding characteristic timescales.
The shorter timescale $t_{\rm steep}$ can be considered to be the responding timescale of the system.
Before this time, the system has not responded significantly to the disturbance.
After the response time, the system undergoes significant changes due to the energy injection, deviating from its equilibrium state.
The larger timescale $t_{\rm flat}$ can be understood as the relax timescale of the system.
At this time, the energy dissipation in the system offsets the energy injection, making the system to return to equilibrium.

The PSD for each source obtained here only displays a relaxation timescale $t_{\rm flat}$.
It could be the result of the fact that the responding timescale is very close to the relaxation timescale in the SHO PSD.
This indicates that once the energy injection changes the state of the system, the energy is immediately released, and the system promptly returns to equilibrium (the physics is also for Mat$\acute{\rm e}$rn$-3/2$ model). That is, the system undergoes an abrupt energy release.

Unlike the long-term variability, the variability patterns observed in the short-term light curves suggest that the extreme outburst may be driven by the process of an abrupt energy release in jets.
Interestingly, the X-ray light curves of some normal X-ray bursts (XRB) in magnetar SGR 1935+2154 can also be described by the SHO kernel approaching the critically damped regime, displaying a PSD shape similar to that of the two $\gamma$-ray flares \citep{2024ApJ...971...26T}.
Magnetar bursts are usually associated with magnetic reconnection \citep[][]{2021ApJ...921...92B,2020ApJ...900L..21Y}.
This indicates that the short-term variability of the 3C 454.3 flare and 3C 279 flare could be driven by magnetic reconnection in jets. The statistic analysis for the $\gamma$-ray flare characteristics of 3C 454.3 \citep{2018RAA....18...40Z}, and the analysis of flare envelope fitting and energy spectra for $\gamma$-ray flares of 3C 279 observed by Fermi-LAT, also support that the magnetic reconnection is the mechanism causing such short-term variability \citep{2020NatCo..11.4176S}. 
The dissipation timescale is more likely to be associated with the size of the magnetic reconnection region \citep{2020MNRAS.494.1817D}, which may be informed by the characteristic timescale we obtained here.

Overall, the three models adopted in our analysis offer valuable insights into both the characterization and interpretation of flare variability, but they impose fixed constraints on the PSD shape (slope).
The stretched exponential model offers greater flexibility by allowing a tunable PSD slope (at high-frequency) in the range between -1 and -3 \citep{2013ApJ...765..106Z,2020ApJ...900..117Z,2025arXiv250516884H}. We would like to consider it as a potential extension for further investigation in future work as it provides a more versatile model for capturing diverse variability patterns in astrophysical light curves.

\section{Summary} \label{sec:summary}
Based on our understanding of the non-thermal radiation variability characteristics of blazars \citep[including the long-term stochastic variability and QPOs;][]{2021ApJ...919...58Z,2022ApJ...930..157Z,2023ApJ...944..103Z},
we further investigate the properties of the extreme $\gamma$-ray flares of 3C 454.3 and 3C 279. The GP method is carried out with SHO, DRW and Mat$\acute{\rm e}$rn$-3/2$ as the kernels for the flaring data.
The short-term variability modes for both 3C 454.3 flare and 3C 279 flare are more inclined towards the over-damped SHO mode (approaching a critically damped mode) 
and Mat$\acute{\rm e}$rn$-3/2$ model.
We propose that the variability of the giant $\gamma$-ray flares may reflect physical processes (in jets) analogous to a critically damped response. In such a process/system, magnetic reconnection may be the energy dissipative mechanism, resulting abrupt energy release.
The derived timescale of the variability may provide an indication of the size of the magnetic reconnection region.

\section*{Data Availability}
The data underlying this article can be provided by the first author\footnote{Email:zhanghy@ynu.edu.cn} upon reasonable request.

\section*{acknowledgments}
We acknowledge the funding support from the National Natural Science Foundation of China (NSFC) under grant No. 12393852, and the support from the Postdoctoral Fellowship Program of China Postdoctoral Science Foundation (CPSF) under Grant Number GZB20230618.

\bibliography{main}
\bibliographystyle{mnras}

\appendix

\section{Additional Figures}
\label{sec:appendix_figs}

\begin{figure}
    \centering
    {\includegraphics[width=1\linewidth]{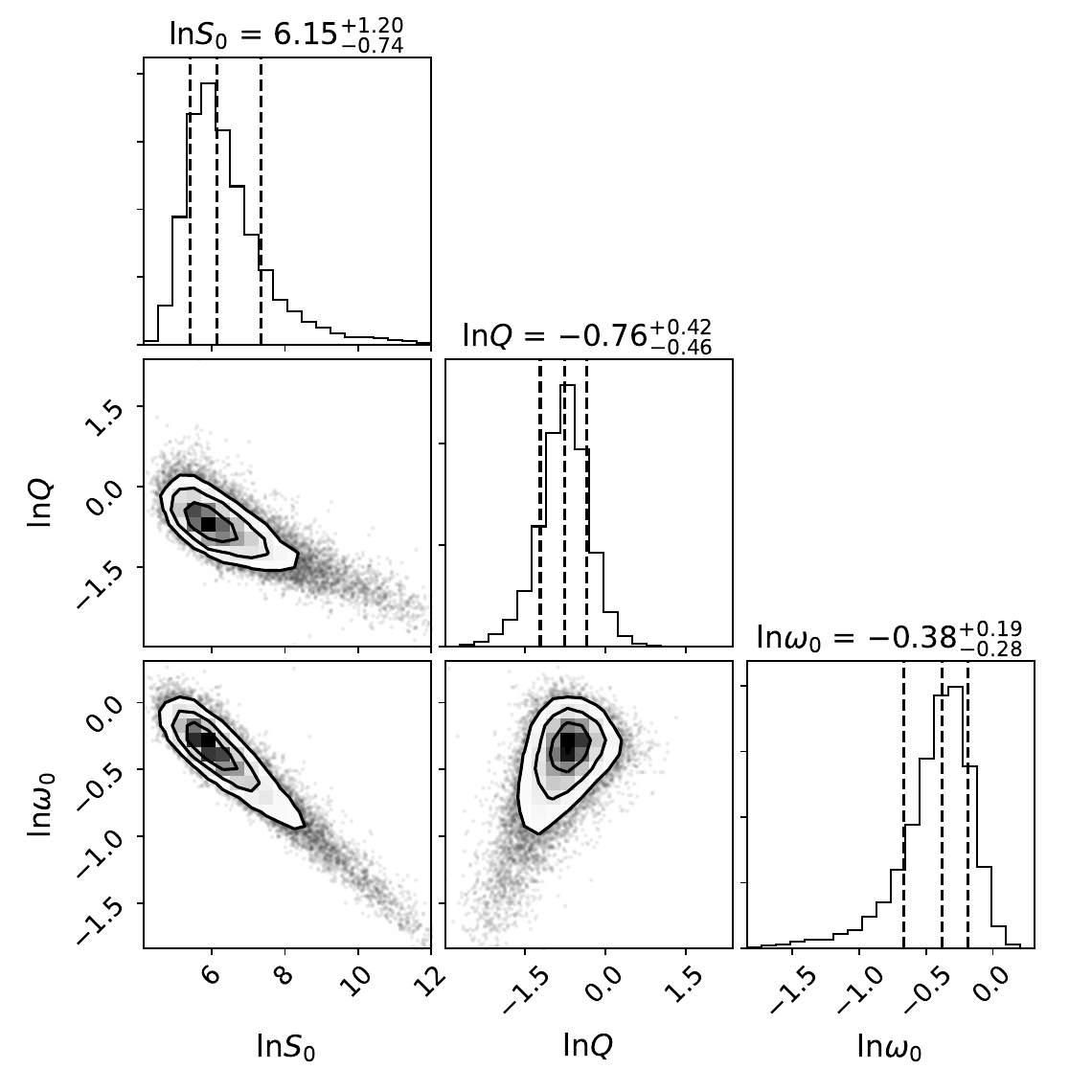}} \\
    {\includegraphics[width=1\linewidth]{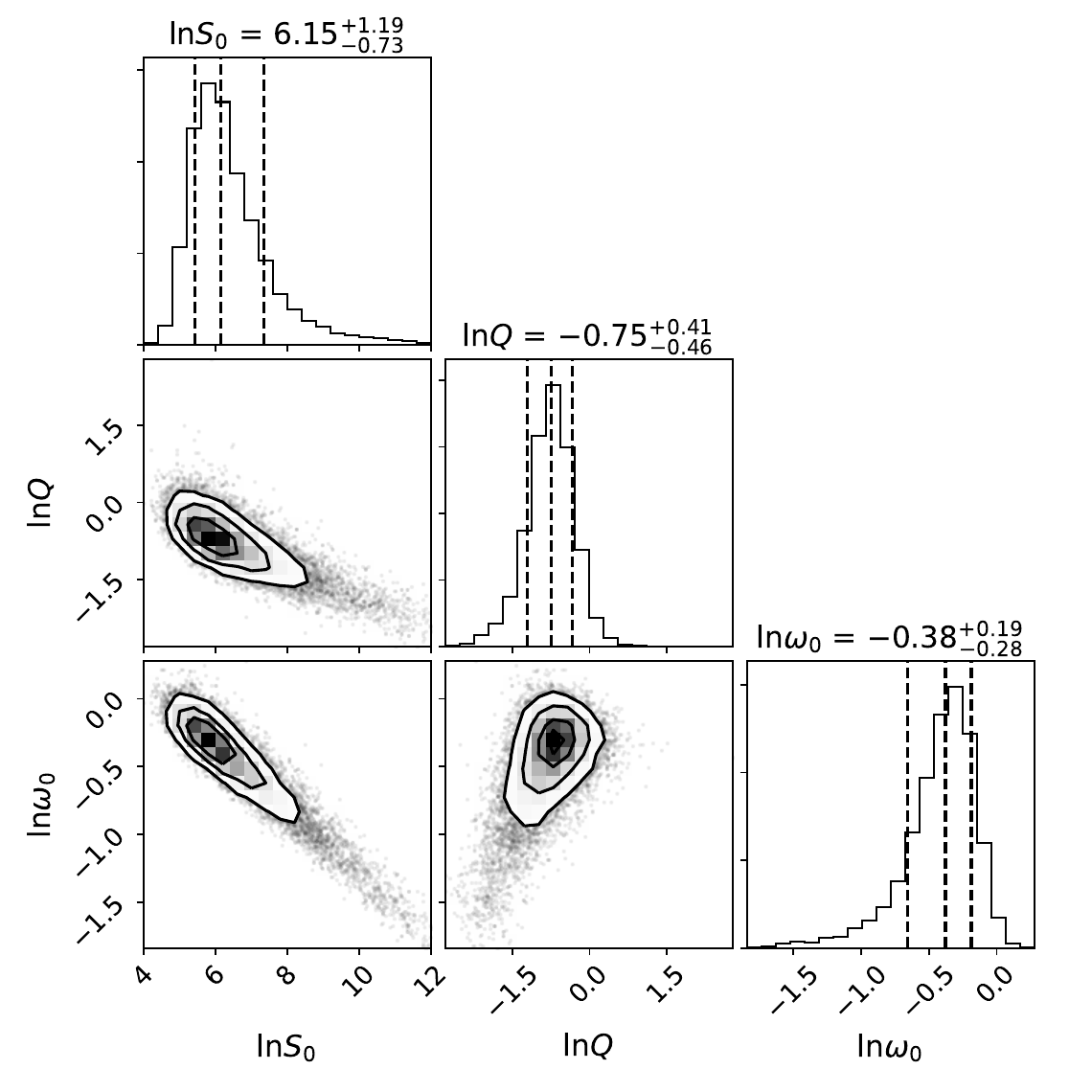}}
    \caption{ The posterior probability density distribution of the SHO parameters for the 3C 454.3 flare. The top panel is obtained from MCMC sampling after thinning, while the bottom panel is from nested sampling.
\label{fig:param-sampling}} 
\end{figure}

We examined the impact of parameter autocorrelation and different sampling methods on inference results. 
As an example, Figure~\ref{fig:param-sampling} shows the posterior distributions of the SHO model parameters fitted to the flare of 3C 454.3, obtained using both thinned MCMC samples and the nested sampling method.

\end{document}